\documentclass{article}

\usepackage{microtype}
\usepackage{graphicx}
\usepackage{subcaption}
\usepackage{booktabs} 

\usepackage{hyperref}

\usepackage[preprint]{icml2026}

\usepackage{amsmath}
\usepackage{amssymb}
\usepackage{mathtools}
\usepackage{amsthm}
\usepackage{physics}

\usepackage[capitalize,noabbrev]{cleveref}

\theoremstyle{plain}

\theoremstyle{definition}

\theoremstyle{remark}

\usepackage[textsize=tiny]{todonotes}

\icmltitlerunning{Submission and Formatting Instructions for ICML 2026}

\begin{document}

\twocolumn[
  \icmltitle{Generalization vs. Memorization in Autoregressive Deep Learning:\\
Or, Examining Temporal Decay of Gradient Coherence}



  \icmlsetsymbol{equal}{*}

  \begin{icmlauthorlist}
      \icmlauthor{James Amarel}{lanl}
      \icmlauthor{Nicolas Hengartner}{lanl}
      \icmlauthor{Robyn Miller}{lanl}
      \icmlauthor{Kamaljeet Singh}{ua}
      \icmlauthor{Siddharth Mansingh}{lanl}
      \icmlauthor{Arvind Mohan}{lanl}
      \icmlauthor{Benjamin Migliori}{lanl}
      \icmlauthor{Emily Casleton}{lanl}
      \icmlauthor{Alexei Skurikhin}{lanl}
      \icmlauthor{Earl Lawrence}{lanl}
      \icmlauthor{Gerd J. Kunde}{lanl}
  \end{icmlauthorlist}

  \icmlaffiliation{lanl}{Los Alamos National Laboratory, Los Alamos, NM 87545}
  \icmlaffiliation{ua}{The University of Arizona, Tucson, AZ, 85721}

  \icmlcorrespondingauthor{James Amarel}{jlamarel@lanl.gov}

  \icmlkeywords{Machine Learning, ICML}

  \vskip 0.3in
]



\printAffiliationsAndNotice{}  

\begin{abstract}
Foundation models trained as autoregressive PDE emulators hold significant
promise for accelerating scientific discovery through their capacity to both
extrapolate beyond training regimes and 
efficiently adapt to downstream tasks despite a paucity of examples for
fine-tuning. However, reliably achieving genuine generalization---a necessary
capability for producing novel scientific insights and robustly performing
during deployment---remains a critical challenge. 
Establishing whether these requirements are met demands 
evaluation metrics capable of clearly distinguishing genuine 
generalization from mere memorization.
We apply the influence function formalism to systematically characterize how
autoregressive PDE emulators assimilate and propagate information derived from
diverse physical scenarios, revealing fundamental limitations of standard
models and training routines in addition to providing actionable insights regarding
the design of improved surrogates.
\end{abstract}

\section{Introduction}
Machine learning surrogate models has emerged as a powerful technique for efficiently
approximating the solutions of computationally intensive partial differential equations
(PDEs).  These emulation methods range from purely data-driven approaches, trained on
high-fidelity simulation data, to physics-informed neural networks,  which integrate PDE
structures directly into the training loss, enforcing physical laws as soft constraints
\cite{Raissi2019PINN}. Such models hold promise for achieving significant computational
acceleration in applications such as fluid dynamics
\cite{takamoto2024pdebenchextensivebenchmarkscientific,
lippe2023pderefinerachievingaccuratelong,
gupta2022multispatiotemporalscalegeneralizedpdemodeling,
ohana2025welllargescalecollectiondiverse, herde2024poseidonefficientfoundationmodels},
climate modeling  \cite{bodnar2024foundationmodelearth}, and materials science
\cite{batatia2024foundationmodelatomisticmaterials}, enabling rapid research and development
across diverse scientific and engineering disciplines. Despite these advances, 
reliable generalization and robustness remains a critical challenge
\cite{krishnapriyan2021characterizingpossiblefailuremodes}. 
Before surrogate models can be safely deployed in operational environments
demanding generalization beyond their training data, it is essential to develop
methods capable of quantifying risk profiles and ensuring trustworthiness of predictions.

Distinguishing between memorization of training examples and genuine
generalization is critical to evaluating model robustness; diagnostic tools such as
influence functions
\cite{koh2020understandingblackboxpredictionsinfluence, bae2022influencefunctionsanswerquestion},
leverage scores, and gradient alignment analyses offer
promising avenues for characterizing this balance, revealing whether models
rely appropriately on generalized understanding or disproportionately on
memorized patterns
\cite{fort2020stiffnessnewperspectivegeneralization,
chatterjee2020coherentgradientsapproachunderstanding,
chatterjee2022generalizationmysterydeeplearning,
zielinski2020weakstronggradientdirections}.

Autoregressive models, useful for their promising extrapolation capabilities,
accumulate errors during inference, in part due to the inevitable distribution shift that originates
with the usage of model outputs as inputs to drive the predicted evolution arbitrarily far
into the future \cite{lee2023autoregressiverenaissancein, brandstetter2023messagepassingneuralpde}.
While such inference-time error accumulation is significant, we reveal a decisive learning
limitation that also contributes to the difficulty of achieving stable long-term rollouts:
gradient signals fail to propagate coherently across time, which implies that ordinary
training lacks a mechanism for generalizing from supervision of one-step predictions to
multi-frame dynamical evolution governed by a shared update structure.
The length of time that a model performs (and is confident) prior to excessive prediction
error defines a ``trust horizon'' for forward prediction 
that is contingent on its encoding of the true data‑generating mechanisms---the physics---rather than merely exploiting empirical correlations among proximal points in feature space
to perform local statistical interpolation. 

Traditional benchmarks, such as point-wise mean squared error
evaluations on limited validation datasets, often
fail to adequately capture surrogate model reliability, especially when faced
with variations in initial or boundary
conditions, mesh resolutions, or varying physical parameter regimes \cite{simshift}. Physics-informed
metrics, including conservation-law violation assessments, PDE residual norms,
analytical-limit checks, and numerical stability evaluations, have been
proposed to better reflect model robustness \cite{Karniadakis2021SciMLReview},
yet even these enriched criteria are not guaranteed to fully quantify the true worst-case
prediction errors. Indeed, empirical accuracy metrics based on finitely many
examples can dramatically underestimate the true worst‐case error, especially
when the data is noisy, sparse, or incompletely understood \cite{vap1998}.

In scientific machine learning, limited availability of high-fidelity simulation data often
results in narrow training distributions, making it challenging to develop robust emulators.
On queries poorly represented by the training set, data-driven predictive models
risk producing non-physical artifacts, such as violations of conservation laws, causality,
or symmetry. While transfer learning and multi-fidelity methods have emerged to alleviate
data scarcity, ensuring physically consistent generalization remains a significant challenge
\cite{herde2024poseidonefficientfoundationmodels}. Towards addressing this gap,  current
research increasingly emphasizes the development of PDE foundation models designed to
achieve robust and unified generalization across diverse physical scenarios
\cite{sun2025foundationmodelpartialdifferential, ye2024pdeformer,
herde2024poseidonefficientfoundationmodels,
subramanian2023foundationmodelsscientificmachine}.
Contemporary PDE emulators employ a variety of architectures \cite{Li2021FNO,
Lu2021DeepONet, gregory2024equivariantgeometricconvolutionsemulation,
shankar2023importanceequivariantinvariantsymmetries}; however, most large-scale deployments
rely on UNet \cite{ronneberger2015unetconvolutionalnetworksbiomedical} or
Transformer backbones \cite{vaswani2023attentionneed,
liu2021swintransformerhierarchicalvision, dosovitskiy2021imageworth16x16words}, and there
remains no consensus on which model variant is most capable at scale.
One
must balance ease of optimization with the incorporation of physics priors, but
quantitative tools for comparing loss-landscape properties across these
architectures remain under-explored. 

Insight into surrogate model behavior beyond static accuracy metrics can be
gained through analysis of the model gradients. Combining test example error evaluation with gradient examination
allows for interpolation of prediction errors across the underlying data
manifold; for instance, PINNs can be
certified with continuous-domain error
bounds \cite{eiras2024efficienterrorcertificationphysicsinformed}.
By quantifying gradient overlap among different training examples, it is
possible to identify potential conflicts or synergies present during learning
and inherent to fully trained models.
Precisely how gradients derived from individual training samples propagate
through model parameters is formalized  through the use of influence functions \cite{Hampel01061974, Cook1980}.
Influence functions were originally developed in robust
statistics \cite{huber2009robust} to quantify how small perturbations of a data point in the
training set affect model parameter estimations \cite{koh2020understandingblackboxpredictionsinfluence, bae2022influencefunctionsanswerquestion}. 
Diagonal elements of the influence function measure each training example’s self-leverage;
high-leverage points thereby identifying data that exerts disproportionate impact during
training. 

For PDE surrogates, the influence framework can also pinpoint examples
providing gradient signals that exacerbate violations of physical constraints
\cite{naujoks2024pinnfluenceinfluencefunctionsphysicsinformed}. Furthermore,
influence functions reveal spatio-temporal correlations inherent in PDE
emulator learning \cite{wang2025gradientalignmentphysicsinformedneural},
distinguishing between memorization and  generalization in cases where the
underlying solution operator lacks explicit space-time dependence, in addition
to exposing gradient misalignments across distinct initial conditions and
inputs that are well separated in feature space. When applied to PDE foundation
models, these techniques systematically characterize model stability,
generalization capability, and uncertainty under structured domain shifts and
multi-physics scenarios, in addition to  uncovering subtle failure modes
typically missed by conventional evaluation metrics, thereby enabling targeted
refinements of model, architecture, and training routines that yield more
robust, physically-consistent, data-driven  models
\cite{ren2019learningreweightexamplesrobust,
zhang2021learningfastsamplereweighting}.

\section{Related Work}

Influence functions are powerful tools for
understanding model behavior and data importance 
\cite{koh2020understandingblackboxpredictionsinfluence, bae2022influencefunctionsanswerquestion}.
Robust and interpretable criteria
for detecting anomalous inputs follow from
techniques
that analyze the alignment of
gradients \cite{wang2025gradientalignmentphysicsinformedneural}
by quantifying directional consistency with in-distribution data
\cite{huang2021importancegradientsdetectingdistributional}, employ
orthogonal projection
\cite{behpour2023gradorthsimpleefficientoutofdistribution} to isolate anomalous
components, and outlier gradient analysis
\cite{chhabra2025outlier}.

Fort et al.
\cite{fort2020stiffnessnewperspectivegeneralization}
define stiffness in terms of
the dot-product between the loss-gradients of two inputs. A positive stiffness then means that
a stochastic gradient descent (SGD) step benefiting one example simultaneously lowers the loss of the other, evidence that
the network assimilated shared, transferable features. Two summary statistics: sign-stiffness
and cosine-stiffness, emphasize inter-class and intra-class correlations, respectively. 
Plotting stiffness against input-space distance yields a dynamic correlation length---the
distance where average stiffness first crosses zero---which shrinks over epochs, revealing how
the learned function becomes progressively more localized as specialization sets in.

The Coherent Gradients Hypothesis \cite{chatterjee2020coherentgradientsapproachunderstanding}
proposed that
per‑example gradients tend to align for similar inputs, so
SGD steps amplify directions supported by many examples while suppressing idiosyncratic
ones, steering the network toward functions that generalize rather than memorize.
Extensions of the Coherent Gradients Hypothesis 
\cite{zielinski2020weakstronggradientdirections}
posit
that SGD updates aligned across multiple training examples (strong directions) underpin
generalization, whereas idiosyncratic updates (weak directions) promote memorization. They
introduce optimizers that suppress weak
directions without computing per-example gradients, dramatically reducing the train-test
gap-even in the presence of heavy label noise-and thereby offer the first large-scale
confirmation of the hypothesis.
Complementing this view, He and Su
\cite{he2020localelasticityneuralnetworks}
establish the notion of local elasticity:
in some neural networks, a parameter update perturbs predictions only within a narrow neighbourhood
around the training point.

PINNfluence
\cite{mlodozeniec2025influencefunctionsscalabledata}
interrogates a trained physics-informed neural network
under perturbations to the PDE parameters and reweighting of 
collocation points.
They distill raw pointwise influences into physically meaningful diagnostics such as the
directional indicator, which measures the fraction of influence that propagates downstream
with the fluid flow.

\section{Our Contributions}

We make three key advances toward principled analysis and validation of PDE emulators:

\begin{enumerate}
  \item \textbf{Time-Aware Analysis of Off‐Diagonal Influence Function Elements:}
        A systematic study off-diagonal influence function elements for PDE surrogate models,
        capable of quantifying training-sample leverage across physical time [see \autoref{fig:summary}]. This diagnostic
        sets standards for identifying the learning of persistent nontrivial
        correlations that extend across temporal horizons, thereby identifying when the
        emulator network has internalized fundamental, time-invariant PDE structures.

\begin{figure}[h] 
    \centering
    \includegraphics[width=0.40\textwidth]{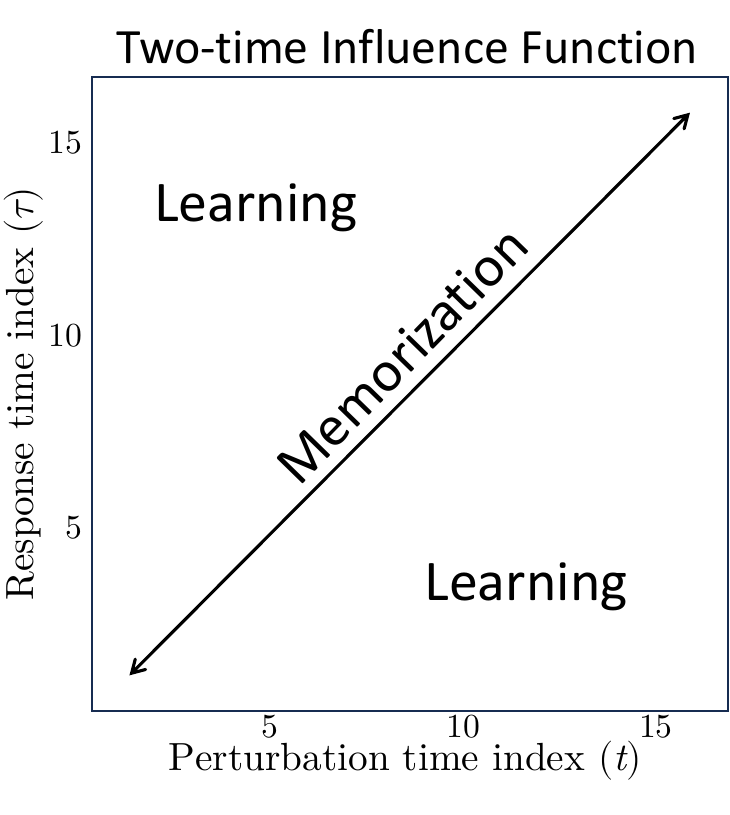}
	\caption{
	Conceptual schematic separating learning from memorization using the
	two-time influence diagnostic. A learning regime corresponds to broad support
	away from the diagonal, meaning that training information at one time affects
	predictions at many other times. A memorization regime corresponds to diagonal
	dominance, meaning that updates are effectively time-indexed and do not encode
	a time-consistent solution operator.
	}
    \label{fig:summary}
\end{figure}

    \item \textbf{Gradient-Coherence Diagnostics Across Initial Condition Classes:}
        We determine the degree of alignment of gradients computed across
        different classes of PDE solutions for two standard architectures, a
        UNet and a ViT. Strong alignment signals the learning of robust, transferable physics, whereas
        weak alignment suggests that the neural network embeds these classes on
        separated regions of the input manifold, with limited gradient coherence,
        despite the fact that the data represents solutions to the same
        underlying PDE. 

	\item \textbf{Dynamic Correlation Length and Curvature Diagnostics:}
		We show that autoregressive PDE emulators generically exhibit a limited dynamic
		correlation length \cite{fort2020stiffnessnewperspectivegeneralization},
		directly observable through the rapid decay of influence with increasing
		feature-space distance. Such feature-space localization provides an explicit,
		training-time explanation for why such models fail to reuse dynamical structure
		or internalize shared physical laws beyond narrow neighborhoods of the data
		manifold.  Complementing this evidence, spectral analyses of the neural tangent
		kernel metric show that low test error is typically achieved in a highly
		anisotropic regime: while most directions remain flat, a small number of
		dominant eigenmodes exhibit large curvature, corresponding to sharp,
		high-sensitivity directions rather than globally robust solutions. This
		spectral imbalance clarifies why apparent interpolation success does not imply
		robustness, and why learned dynamics fail to transfer coherently across time or
		conditions despite favorable one-step performance
		\cite{karakida2019universalstatisticsfisherinformation, anonymous2025measuring}.

\end{enumerate}

This paper is organized as follows. For the readers' convenience, we first
present our central results \autoref{sec:results}, exposing pronounced lack of
generalization capabilities in autoregressive PDE emulators. Technical
details---those covering both the mathematical formulation of the
influence-function framework in addition to our training 
procedures---are provided in \autoref{sec:response} and \autoref{sec:train},
respectively.

\section{Results}
\label{sec:results}
We examine how training information propagates across time and
initial-condition classes in autoregressive PDE emulators, using
influence-based diagnostics evaluated on held-out test data. Across
architectures, physical observables, and datasets, we find that gradient
responses are strongly localized in both time and class, with off-diagonal
influence rapidly decaying, indicating that these models primarily learn time-
and class-indexed update rules rather than a globally consistent dynamical
operator.

\begin{figure}[h] 
    \centering
    \includegraphics[width=0.475\textwidth]{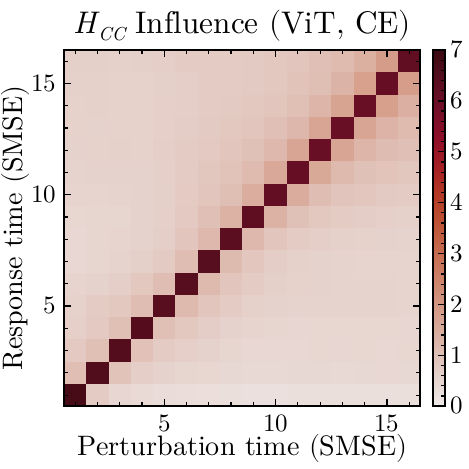}
	\caption{
		Heatmap of two-time influence for our ViTs trained on 
		CE data, shown as a function of perturbation time (horizontal axis) and
		response time (vertical axis). 
        Each pixel reports the intra-class averaged
		response induced by test example gradients
		at the perturbation time. 
        A narrow diagonal ridge corresponds to time-local
		sensitivity consistent with interpolation, rather than generalization;
		substantial off-diagonal structure would indicate time-transferable
		learning.
        For the analogous plot using our UNets, see \autoref{fig:heatmap_UC}. For NS data
        counterparts, see \autoref{fig:heatmap_VN} and \autoref{fig:heatmap_UN}.
    }
    \label{fig:heatmap_VC}
\end{figure}

\begin{figure}[h] 
    \centering
    \includegraphics[width=0.475\textwidth]{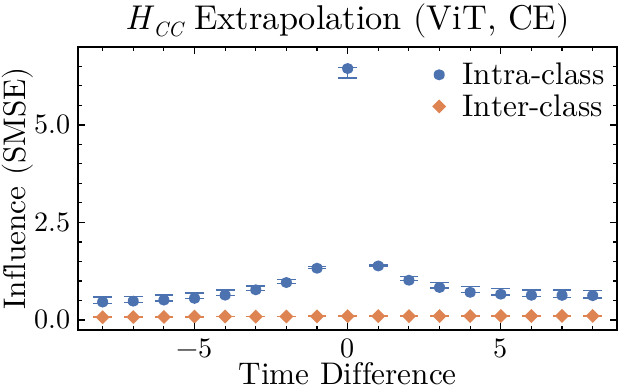}
	\caption{
	Time-lag summary of temporal transferability for our ViTs on CE data. The
	influence is averaged over all time-pairs of the same
	time difference and then split into intra-class pairs (gradient and
	response drawn from the same initial-condition class) and inter-class pairs
	(distinct initial-condition classes). Strong concentration near zero
	time difference indicates that gradient information fails to propagate
	coherently across time, while the lack off inter-class influence
	reveals an absence of physics consistent generalization. 	
    For the analogous plot using our UNets, see \autoref{fig:horizon_UC}. For NS data
    counterparts, see \autoref{fig:horizon_VN} and \autoref{fig:horizon_UN}.
}
    \label{fig:horizon_VC}
\end{figure}

\begin{figure}[h] 
    \centering
    \includegraphics[width=0.475\textwidth]{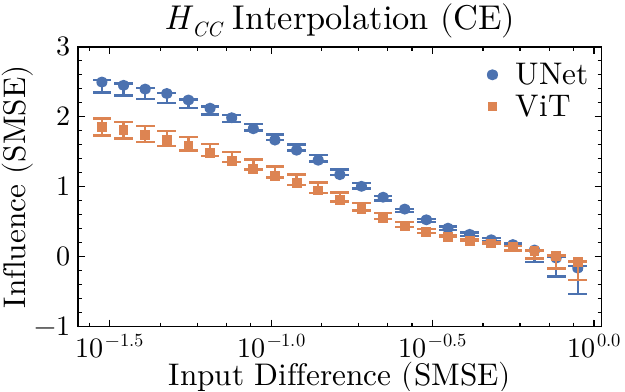}
	\caption{
	Curve fit of the influence as a function of
	feature-space separation between input states for CE data, comparing UNet
	and ViT; rangebars show uncertainty across seeds. A steep decay indicates
	short-range locality on the learned data manifold, implying that parameter
	updates affect only nearby states and generalization is limited.
    For NS data counterpart, see \autoref{fig:rkhs_NS}.
	}
    \label{fig:rkhs_CE}
\end{figure}

Test-data measurements of the two-time influence function [see
\autoref{eq:H_resp}] for both a UNet and a ViT tasked with emulating fluid flow
exhibit rapid temporal decay in the off-diagonal terms [see
\autoref{fig:heatmap_VC}], which indicates that surrogate training constructs
localized vector fields suitable only for interpolation within small
neighborhoods of the training data sub-manifold, rather than the universally
consistent function that is desired based on expectations stemming from our
knowledge of the underlying governing equations. If such models were truly
learning the solution operator to a PDE that lacks explicit time dependence,
gradients derived from examples at a given time would necessarily have a
profound effect on the predictions at any other time, for we know that the true
solution operator must be time-translation equivariant, taking the same
functional form at every point in phase space.  This superfluous time awareness
presents across the entire training trajectory, demonstrating that our models
did not learn the underlying solution operator.
\begin{figure}[h] 
    \centering
    \includegraphics[width=0.475\textwidth]{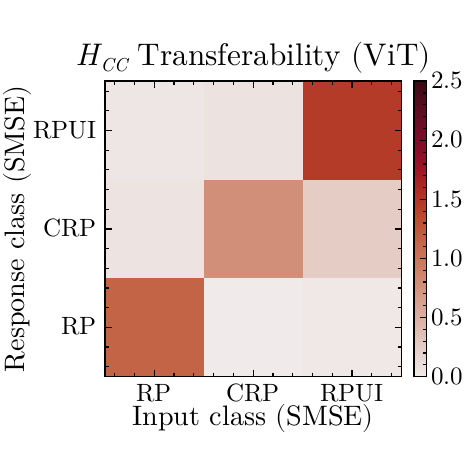}
	\caption{
	Class-to-class transferability matrix for our ViTs trained on the three-class CE
	split labeled RP, CRP, and RPUI. 
    Each entry reports the time-averaged 
	influence of test examples from the
	input class (horizontal axis) on examples from the
	response class (vertical axis). 
    Diagonal dominance indicates class-locked
	gradient geometry; substantial off-diagonal values would imply reuse
	of dynamical features across classes.
    For the analogous plot using our UNets, see \autoref{fig:diag_UC}. For NS data
    counterparts, see \autoref{fig:diag_VN} and \autoref{fig:diag_UN}.
	}
    \label{fig:diag_VC}
\end{figure}
Consistent with the two-time influence maps, the class-to-class transferability matrix in
\autoref{fig:diag_VC} is strongly diagonal, indicating that gradient geometry is effectively class-locked:
updates supported by one initial-condition family produce negligible response in the
others. Furthermore, there is a near-total absence of inter-class influence [see \autoref{fig:horizon_VC}].
The degree of gradient alignment across examples also affords conclusions about the data
manifold sparsity: while all inputs to the network are intimately related as unique
solutions to a shared equation of motion under different initial conditions,  both our ViTs
and our UNets render inputs well separated in the sense that their gradients don't
meaningfully overlap unless their feature space distance  small [see \autoref{fig:rkhs_CE}],
which implies limited generalization over dynamical structure away from nearby states.
Such results challenge a fundamental assumption motivating the development of PDE foundation
models, as it demonstrates that these models are prone to effectively treating different
flow fields as distinct, isolated learning tasks. That this happens even when
said classes of solutions arise merely from different initial conditions to the same
physical process underscores the need for inductive biases to be explicitly incorporated 
during model development. 

\begin{figure}[h] 
    \centering
    \includegraphics[width=0.475\textwidth]{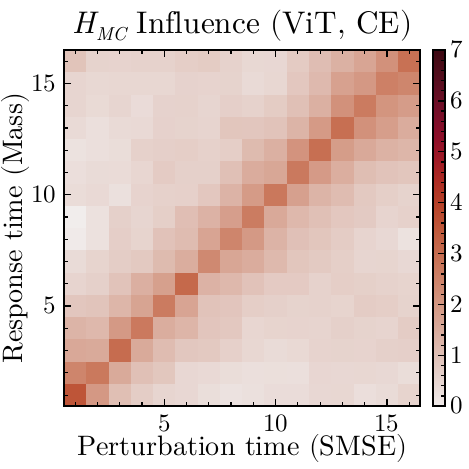}
	\caption{
	Two-time influence map for our ViTs on CE data when the
	response observable is the global mass-consistency signal. The horizontal
	axis indexes the time at which a test perturbation is applied, and the
	vertical axis indexes the time at which the mass-based response is
	evaluated. Off-diagonal support indicates that intra-class mass-related
	gradient information couples distant times.
    For the analogous plot using our UNets, see \autoref{fig:heatmap_UC_mass}. 
    For plots concerning energy conservation, see
    \autoref{fig:heatmap_VC_energy} and
    \autoref{fig:heatmap_UC_energy}.
	}
    \label{fig:heatmap_VC_mass}
\end{figure}
\begin{figure}[h] 
    \centering
    \includegraphics[width=0.475\textwidth]{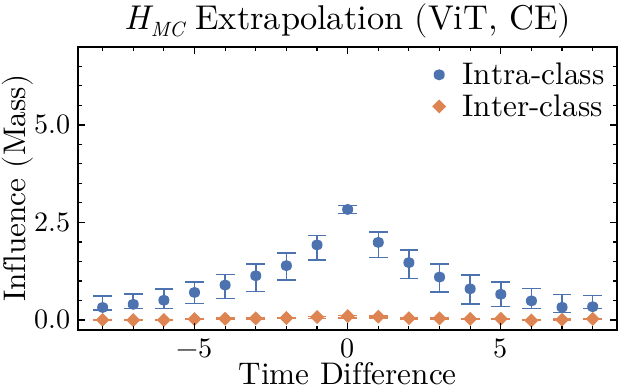}
	\caption{
	Time-lag transferability curve for our ViTs on CE data using the global
	mass-consistency observable. Intra-class averages measure how mass-based
	influence response align within an initial-condition class, while
	inter-class averages measure cross-class reuse of mass-related gradient
	directions. The decay pattern diagnoses whether conservation constraints
	induce transferable structure or remain class-locked.
    For the analogous plot using our UNets, see \autoref{fig:horizon_UC_mass}. 
    For plots concerning energy conservation, see
    \autoref{fig:horizon_VC_energy} and
    \autoref{fig:horizon_UC_energy}.
    }
    \label{fig:horizon_VC_mass}
\end{figure}

In addition to the overlap of cost function gradients, we also considered gradients derived
from physics informed loss functions, such as global mass conservation [see
\autoref{fig:heatmap_VC_mass}] and global energy conservation.  
In all cases, we
observed that the response function decayed off the time-diagonal and was dominated by
intra-class matrix elements [see \autoref{fig:horizon_VC_mass}].  Hence, we conclude that predictive models lacking explicit
inductive biases are not internalizing a unified governing law, but merely allocating
parameters tasked specifically with evolving states associated with a given time along a
given class of trajectories. 

\begin{figure}[h] 
    \centering
    \includegraphics[width=0.475\textwidth]{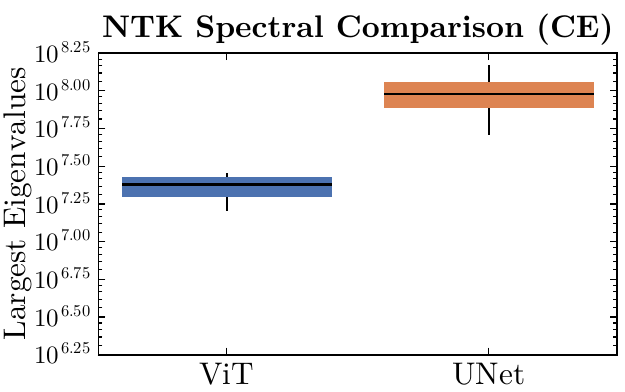}
	\caption{
	Spectral comparison of dominant neural tangent kernel eigenvalues for our UNets
	and our ViTs on CE data. The plotted
	distributions summarize the leading eigenvalue statistics across seeds and
	batches of the trained models. Larger dominant eigenvalues indicate a locally
	stiffer, sharper response geometry.
    For NS data counterpart, see \autoref{fig:eig_NS}.
	}
    \label{fig:eig_CE}
\end{figure}

Lastly, \autoref{fig:eig_CE} shows that the dominant NTK eigenmodes are large, revealing a
stiff, highly anisotropic local response geometry; in particular, low test error coexists
with sharp high-curvature modes rather than a uniformly flat, robust geometry.

\section{Proximal Response Function}
\label{sec:response}
We develop
the influence function as follows,
taking inspiration from
\cite{bae2022influencefunctionsanswerquestion}.
Let \(\theta\) be the current parameter values and
consider the optimization step 
\(
    \theta \leftarrow \theta + \delta \theta,
\)
where the tangent-space displacement \(\delta \theta\) 
minimizes the proximal objective
\begin{equation}
    S(\delta\theta) 
    = dC[\delta \theta]
    + \frac{1}{2}||\hat{y}(\theta) - \hat{y}(\theta + \delta \theta)||_{L_2}^2,
    \label{eq:action}
\end{equation}
with \(\hat{y}\) a neural network.
The stationarity condition \(dS \stackrel{!}{=} 0\)
is satisfied (to linear order) by
\begin{equation}
    \delta \theta^{\mu} = -\,\eta^{\mu \nu}\partial_{\nu} C,
\end{equation}
where
\(
    \eta_{\mu \nu} = J^n_{\mu} J^n_{\nu}
\)
is the neural tangent kernel metric,
\(
J_{\mu}^n = \partial_{\mu}\hat{y}^n
\) 
is the model Jacobian,
and \( n \) indexes a given mini-batch example.
By convention, the components of $\eta$ carry lowered indices, $\eta_{\mu\nu}$,
while those of \(\eta^{-1}\) carry raised indices, $\eta^{\mu\nu}$, i.e. 
\(
\eta^{\mu\alpha}\eta_{\alpha\nu} = \delta^{\mu}_{\ \nu};
\)
\(\eta\) provides the canonical correspondence between covariant and contravariant
components \cite{absil2008}.
\autoref{eq:action} balances the force term $dC$ against the kinetic cost of the update
distance in the \(\eta\) prescribed geometry. Convexity of $\eta$, together
with mild regularity requirements on $C$, guarantees a unique stationary point
of each proximal subproblem. Proximal gradient descent iterates these
subproblems to accumulate a sequence of locally improving displacements that
drives descent of cost function $C$.  
The inverse susceptibility
tensor $\eta^{-1}$ serves as a generalized stiffness operator by propagating gradient
signals to parameter displacements \cite{fort2020stiffnessnewperspectivegeneralization}.

Classical influence functions can be expressed as the Lie derivative of a scalar; they're capable of
probing local gradient coherence, generalization capabilities, and adversarial sensitivity, in
addition to enabling the identification of high-leverage examples. Consider a  scalar
observable \(Q\) and a vector field \(V=-\eta^{-1}(dC)\) derived from the proximal objective. The Lie
derivative of \(Q\)  along \(V\) is
\begin{equation}
    \begin{aligned}
    \mathcal{L}_V Q 
         &= \left(\partial_{\mu} Q\right) \,\eta^{\mu \nu}\, \left(-\partial_{\nu} C\right)\\
         &=  -\left(\frac{\delta Q}{\delta \hat{y}^n},\Pi^{nm}\frac{\delta C}{\delta \hat{y}^m}\right),
    \end{aligned}
    \label{eq:lie_scalar}
\end{equation}
where
\begin{equation}
    \Pi^{nm} = J_{\mu}^n\eta^{\mu \nu}J_{\nu}^m,
\end{equation}
and the inner product \(\left(\cdot, \cdot\right)\) is performed
over feature indices.
Hence, when \(Q\) is a loss function, \autoref{eq:lie_scalar} reduces to the familiar form
of an influence function in  deep learning: the infinitesimal response of the loss,
expressible as a metric-weighted gradient overlap. Likewise, when \(Q\) denotes a model
response and \(V\)  encodes the perturbation to the gradient signal induced by a deformation
of the  input, \autoref{eq:lie_scalar} reproduces the classical influence-function 
expression from robust statistics [see \autoref{app:hat}].

Evidently, the Lie-derivative formulation
of response is well defined at any point along the training trajectory, as it
depends only on the instantaneous training-flow vector field and the induced
local geometry; hence, influence-function analysis of neural networks does not
require attainment of a stationary point to expand about. This perspective
elevates influence from a static sensitivity relevant only near convergence to
a dynamical linear response observable defined throughout optimization.

In the limit of vanishing regularization, \(\lambda \rightarrow 0\), \(\Pi\) becomes
idempotent, assuming the form of a classical hat matrix.  We thus take the view that
diagonal elements reflect statistical leverage, quantifying the self-influence of individual
training examples, while off-diagonals measure cross-influence, i.e. influence between
distinct examples.  High leverage scores identify regions of parameter space with strong
local curvature or limited redundancy, i.e., points with disproportionately large influence
on the global response structure.  Furthermore, the response matrix encodes the pairwise
overlap of example gradients-effectively probing the local loss landscape by highlighting
directions of correlated curvature and shared descent paths. Physical considerations that
guide expectations for the structure of \( \Pi \) are evident on recognizing that we have so
far suppressed feature indices in the model Jacobian. The response matrix \(\Pi^{nm}\)
tracks how gradients derived from each output feature of prediction \(m\) influence each
output feature of prediction \(n\), offering an investigative level of detail across
spacetime, channel, and class dimensions that remains unexplored.  We emphasize that the
proximal penalty in \autoref{eq:action} sets the geometry of the update and clearly
identifies \(\Pi\) as the primary object governing to what extent an infinitesimal
perturbation in the cost function propagates to an observable, such as the test error or
physical consistency of predictions.  To avoid materializing \(\Pi\), which has
\((128\times128\times4\times48)^2 \) elements, we consider macroscopic observables: SMSE, in
addition to global mass and energy conservation.

\subsection{Observables}
Our probe of generalization capabilities proceeds by quantifying the coherence of
gradients derived from test data cost functions of physical and statistical significance.
We introduce three generalized residuals
\begin{subequations}
    \begin{align}
        r_{\text{C}} &= \frac{\delta C_{\text{SMSE}}}{\delta \hat{y}_{\theta}} = \frac{1}{4}\sum_{c}\frac{\hat{y}^c_{\theta} - y^c}{\text{RMS}(y^c)},\\
        r_{\text{M}} &= \frac{\delta C_{\text{Mass}}}{\delta \hat{y}_{\theta}} = \frac{M[\hat{y}_{\theta}] - M[x]}{M[x]},\\
        r_{\text{E}} &= \frac{\delta C_{\text{Energy}}}{\delta \hat{y}_{\theta}} = \frac{E[\hat{y}_{\theta}] - E[x]}{E[x]},
    \end{align}
\end{subequations}
where \( M \) (\(E\)) computes the total mass (energy) of its argument,
\( x \) is the input state that evolves to \( y \), i.e. \( y = U[x] \),
where \( U \) is defined in \autoref{eq:U}, and 
\( \hat{y} \) is the neural network approximation to \( y \). 
Recall that each training example is comprised of pairs
\(  (x,y)\;=\;(s_t^{n},\,s_{t+1}^{n}) \)
of states \(s\) sharing a common initial configuration indexed by \(n\), and
related by the compressible Euler evolution operator.

Viewed as a coupling matrix over residuals, the diagonal blocks of \(\Pi\)
recover the usual influence (e.g., how a perturbation in the SMSE affects SMSE itself),
while the off-diagonal blocks encode cross-coupling, quantifying how a change in the SMSE
residual at one time step or sample is converted into the conservation residual at another
time step or sample, and vice versa. 

It is useful to introduce
following notation for the remaining external indices of the response matrix
\begin{equation}
    H_{AB}(t, n | \tau, m) = \left(r_A^{nt}, \Pi_{t\tau}^{nm} r_B^{m \tau}\right),
    \label{eq:H_resp}
\end{equation}
where \(n, m\) index trajectories, defined by distinct initial conditions; indices \(t\) and  \(\tau \)
specify the time step along said trajectories. \(H_{CC}\) gives the change in SMSE due to an
SMSE perturbation,  while \(H_{MC}\) and \(H_{EC}\) propagate the effect of gradients
derived from SMSE  into the physics informed and mass and energy conservation errors,
respectively. 

We report influence in a standardized form by normalizing with respect to the empirical
variance of perturbations within each mini-batch, so that the baseline model-corresponding
to unstructured stochastic variability-naturally sets the reference scale to unity. In this
normalization, departures from one directly indicate influence beyond what is expected from
random mini-batch fluctuations, providing a principled scale for interpreting both amplified
self-responses and suppressed cross-responses \cite{heritier1994robust,lu1997standardized}.
Matrix elements of \(\Pi\) were determined for six different mini-batches, each
of which contained three trajectories corresponding to distinct initial
conditions, for each seed of each model architecture trained, across two datasets [see
\autoref{sec:chi}].
\section{Data and Training}
\label{sec:train}

We trained neural network surrogate models to approximate the evolution of
two-dimensional compressible Euler flows, provided by the PDEGym dataset
\cite{herde2024poseidonefficientfoundationmodels}. 
Specifically, we used a dataset that contains 
three classes of initial conditions, namely, the four quadrant Riemann problem
with (CE-RPUI) and without (CE-RP) uncertain interfaces, in addition to the
curved Riemann problem (CE-CRP). This data is particularly valuable for
studying the progression from a linear wave regime with discontinuities to fully
developed turbulence, a crossover that poses computational and
analytical challenges due to the presence of sharp wave-fronts and emergent
nonlinear interactions. While the CE flows
exhibit comparable large-scale structures, they also display qualitatively distinct
behaviors. In particular, CE-RPUI initial configurations
give rise to complex finger-like instabilities in the flow field that are
absent or less pronounced in both CE-RP and CE-CRP. 
In total, we used \(6,500\) trajectories 
for each of the three classes of initial conditions;
for each trajectory, we used the first \(16\) time steps, for
total of approximately \(110,000\) training pairs requiring greater than \(150\) GB memory.

Since instantaneous flow states alone cannot distinguish viscous Navier-Stokes
flows from their inviscid Euler counterparts, we do not combine compressible
Euler data with Navier-Stokes data. Furthermore, rather than representing a
fluid state using the velocities and pressure in addition to density, as was
done by Poseidon \cite{herde2024poseidonefficientfoundationmodels}, we used the
momentum and energy fields; we expect that this setup will better facilitate 
the learning of all four conservation laws.

Each snapshot of the flow state at discrete time $t$ is represented as a set of
spatially discretized fields $\rho_{\text{mass}}, \rho^i_{\text{mom}},
\rho_{\text{energy}}$ on a uniform grid
of size \(128 \times 128\), where $\rho_{\text{mass}}$ denotes
mass density, $\rho^i_{\text{mom}}$ are the Cartesian components of momentum
density, and $\rho_{\text{energy}}$ is energy density. The model \(
\hat{y}_{\theta} \) is trained to emulate the compressible Euler evolution,
i.e., \( \hat{y}_{\theta} \approx U \), where the operator \( U \) enacts
\begin{equation}
    s_{t+1} \;=\; U[s_{t}],
    \label{eq:U}
\end{equation}
with $s_{t}$ the collection of state variables at timestep $t$, via
optimization of the weights
\(\theta \).
Specifically, we used 
the Adam optimizer \cite{kingma2017adammethodstochasticoptimization}
with learning rate \(5\times 10^{-4}\) and weight decay 
\(\lambda = 10^{-4}\)
to minimize a scaled mean squared error (SMSE) between predicted and true states 
\begin{equation}
    \label{eq:smse}
    C_{\text{SMSE}}(\theta) = \frac{1}{N}\sum_{n=1}^{N}\frac{1}{4}\sum_c \frac{\vert\vert\hat{y}^c_{\theta}(s_{t_n}) - s^c_{t_n+1}\vert\vert_{L_2}^2}{\text{RMS}(s^c_{t_n+1})},
\end{equation}
on mini-batches containing \(N = 48 \)  transitions \(s_t \to s_{t+1}\), chosen randomly
from the training set; here, the \(L_2\) norm is computed over the spatial degrees of
freedom and the channel index \( c \) runs through mass, both cartesian components of the
momentum, and energy, respectively. \(\mathrm{RMS}(s^c)\) is computed channel‐wise by taking the spatial
root‐mean‐square of \(s^c\). Thus, \autoref{eq:smse} strikes a balance between ordinary and
relative mean‐squared‐error; normalizing each channel’s squared error by the target fields'
characteristic amplitude favors examples containing pronounced features, 
but does not completely drown out gradients derived from  relatively quiescent flows,
thereby facilitating accurate capture of high-energy shocks and wavefronts without harsh
under-emphasis of small-amplitude features. Moreover, the scaling of \autoref{eq:smse}
renders dimensionless the matrix elements of interest to this work.

In order to compare the results of our experiments across model architectures,
we trained both a UNet
\cite{ronneberger2015unetconvolutionalnetworksbiomedical} and a vision
transformer (ViT) \cite{vaswani2023attentionneed,
liu2021swintransformerhierarchicalvision, 
dosovitskiy2021imageworth16x16words}. Our UNet was based on BigGAN
\cite{brock2019largescalegantraining}, with four down-sampling blocks and
\(24\) channels after the initial embedding layer, for a total of about
\(13\)-million parameters. Our vision transformer was of layer-depth six, with
\(256\) channels, for a total of about \(5\)-million parameters, fewer than our
UNet due to memory constraints on the \(40\) GB A100s that we used for
training. 
To supplement our compressible Euler study, we repeat our analysis for
velocity fields corresponding to solutions of the Navier-Stokes equations with NS-BB,
NS-Gauss, and NS-Sines initial conditions, which presents a distinct feature space and flow
morphology, involving smoother, viscosity-regularized transport with vorticity-dominated
structure [see \autoref{fig:opt_curve_NS}].

\begin{figure}[h] 
    \centering
    \includegraphics[width=0.475\textwidth]{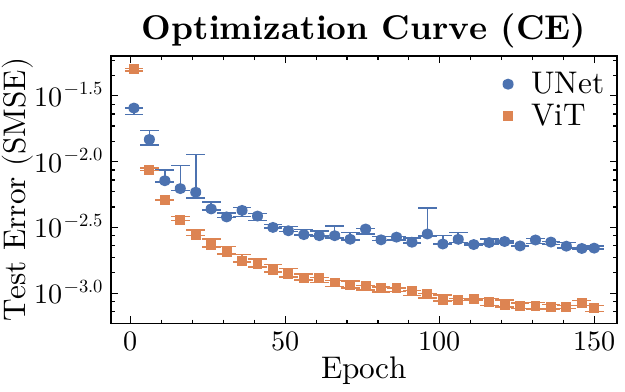} 
	\caption{
		Optimization trajectories for the CE task, shown as
		test scaled mean squared error (SMSE) versus training epoch for our UNets and
		our ViTs. Markers denote the median model performance, while vertical range
		bars indicate the variance across model seeds at each epoch. Training is halted
		in the late-stage learning regime to test whether the induced local gradient
		geometry is sufficient to identify a physically consistent, generalizing
		solution.
        For the corresponding NS-data plot, see \autoref{fig:opt_curve_NS}.
        For representative rollout predictions, see \autoref{fig:rollout_ViT_CE}
        and \autoref{fig:rollout_UNet_CE}.
	}
    \label{fig:opt_curve_CE}
\end{figure}

The validation losses for each of our CE-tasked models is shown in \autoref{fig:opt_curve_CE}. Despite
possessing fewer parameters, our ViT model consistently outperformed the UNet. Each of the
two architectures was trained three times, sharing those three seeds that controlled
initialization and dataset split. 
The training of each model was performed in distributed mode across
two such A100s.

\section{Conclusion}
Our research reveals a critical shortcoming common to physics-agnostic PDE emulators: it
is not an immediate consequence of large-scale multi-scenario training that the resulting trained
model can satisfy those stringent expectations that follow from the governing equations 
one is trying to emulate.
Physical expectations demand a shared feature basis across trajectories, yet
our results reveal a failure of both UNets and ViTs to support
a nontrivial off-diagonal response.
This mismatch underscores the importance of enforcing principled
physics-based constraints, either as weak regularizers during training, or baked in strictly
through architectural design. 
By measuring gradient overlap between classes of initial conditions we reveal an absence
of coherent gradients, which suggests limited learning of robust, transferable physics.
This demonstrates that both ViT and UNet surrogates embed these solution classes on
nearly disjoint manifolds, challenging the efficacy of current multi‑scenario training
pipelines.

We demonstrate that influence functions form a versatile diagnostic framework
and demonstrate their effectiveness in revealing the degree of balance between
memorization and generalization in autoregressive predictors.  This analysis
suggests that ordinary data-driven PDE emulators behave as statistical
estimators, producing predictions primarily based on those training examples
that lie within a neighborhood of the input query. While this localized
learning mechanism provides resilience against noisy data, it also restricts
generalization, and indicates that the learned data manifold geometry is
composed of largely isolated regions.

In summary, we highlight a new, concrete, and targetable characteristics---time
and class aware cross-influence---to guide researchers in designing algorithms
capable of learning the underlying generative process and achieving reliable
long-term rollouts.
\bigskip
\newpage
\noindent

\section{Electronic Submission}


\section*{Software and Data}
We trained our models on
the openly available dataset PDEGym
\cite{herde2024poseidonefficientfoundationmodels}   using Lux.jl
\cite{pal2023lux, pal2023efficient}, with Zygote.jl as our auto-differentiation
backend \cite{Zygote.jl-2018}. Plots in this manuscript were generated using
Makie.jl \cite{DanischKrumbiegel2021}.

The code used in this work is publicly available at 
\href{https://github.com/lanl/PDEHats}{https://github.com/lanl/PDEHats}.
Additionally, trained models and gradient data are available from the authors upon reasonable
request.

\section*{Acknowledgements}
Research presented in this report was supported by the
Laboratory Directed Research and Development program of Los Alamos National
Laboratory under project number(s) 20250637DI, 20250638DI, and 20250639DI.
This research used resources provided by the Los Alamos National Laboratory
Institutional Computing Program, which is supported by the U.S. Department of
Energy National Nuclear Security Administration under Contract No.
89233218CNA000001. It is published under LA-UR-25-28084.


\section*{Impact Statement}
This work contributes a diagnostic framework for distinguishing mere memorization from genuine
generalization in autoregressive surrogate models, with particular relevance to scientific
and engineering applications where model failures can have downstream consequences. By
exposing training-dynamics limitations that are not visible through standard accuracy
metrics, our analysis helps identify when learned surrogates are likely to be brittle under
long-horizon rollout or distribution shift, informing safer deployment in settings such as
climate modeling, fluid dynamics, and materials simulation. The methods introduced here are
diagnostic rather than prescriptive and do not directly enable new capabilities for misuse;
instead, they promote transparency and reliability by clarifying when and why models fail to
internalize shared physical structure. More broadly, our research encourages the development of
learning algorithms and evaluation practices that prioritize robustness and
interpretability, supporting the responsible use of machine learning in high-consequence
scientific workflows.

\bibliography{manu}
\bibliographystyle{icml2026}
\bigskip
\newpage
\appendix
\section{Hat Matrix}
\label{app:hat}

To obtain the related expression familiar from classical influence function theory, 
let \(Q = \hat{y}\) and
\begin{equation}
    \delta C = \frac{\delta C}{\delta y}\delta y,
    \label{eq:modify}
\end{equation}
where \(\delta y\) is a target feature variation.
Then
\begin{equation}
    \delta \hat{y}^n = \mathcal{L}_V \hat{y}^n 
=- \Pi^{nl}\frac{\delta^2 C}{\delta \hat{y}^l \delta y^m} \delta y^m;
\end{equation}
when \(C \) is the mean squared error cost,
\(
    {\delta \hat{y}}/{\delta y} = \Pi.
\)
\autoref{eq:modify}
allows for investigating the influence of both physics-informed and numerical-routine aware data modifications, 
which we save for future work.

\section{Determination of \(\eta^{-1}\)}  
\label{sec:chi}

In practice, we include an $\ell_{2}$-type regularizer corresponding to
the weight-decay term used in AdamW driven training
\cite{kingma2017adammethodstochasticoptimization, deW}. Although this penalty is not intrinsic to
the model geometry---being defined with respect the ambient Euclidean coordinates---it
remains a useful extrinsic regularizer when viewing the parameter manifold as embedded in a
product of real coordinate spaces. Concretely, we consider the regularized metric
\begin{equation}
    \eta_{\mu\nu} \rightarrow
    \eta_{\mu\nu} + \lambda\,\delta_{\nu\mu} ,
    \label{eq:chi}
\end{equation}
with weight decay $\lambda$ providing mass to the zero modes of
$\eta$, thereby weakly lifting its flat directions. 

\begin{figure}[h] 
    \centering
    \includegraphics[width=0.475\textwidth]{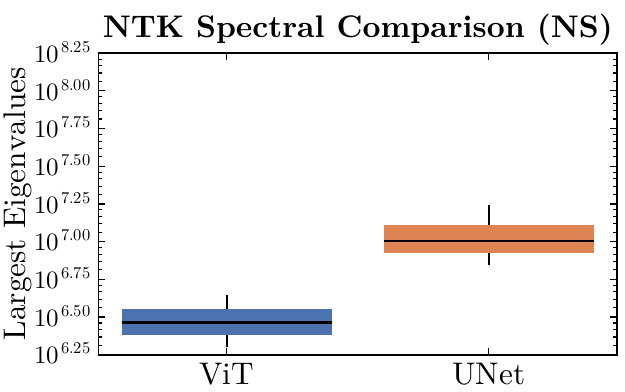}
	\caption{
        Same as \autoref{fig:eig_CE}, except for NS data.
	}
    \label{fig:eig_NS}
\end{figure}

We apply
an iterative matrix-free solver, specifically the CRAIG method \cite{craig} provided by the
Krylov.jl package \cite{montoison-orban-2023}, which is formally equivalent to conjugate
gradient descent, to efficiently approximate the required sensitivities.  
A direct inversion to determine \(\eta\) is not computationally feasible because of
the large number of trainable parameters in our models.
While this
approach does not leverage commonly used scalable approximations \cite{george_nngeometry,
TransferLab_team_pyDVL_2024}, such approximations do not provide error control. When evaluating our UNet models, we determine the action of \(
\eta \) with a relative error tolerance of \(1.5\times10^{-2}\). We reach similar absolute error for our ViT models on using a
relative tolerance of \(5 \times 10^{-2}\); our ViTs have both fewer parameters 
and smaller dominant NTK eigenvales than our UNets [see \autoref{fig:eig_CE} and \autoref{fig:eig_NS}].  

\section{Compressible Euler}
\label{app:euler}
The compressible Euler equations in two-spatial dimensions can be expressed in terms of four
continuity equations, each of which is of the form
\begin{equation}
    \partial_t \rho_{c} +\ \nabla \cdot \mathbf{J}_{c} = 0, 
    \label{eq:cont_eq}
\end{equation}
where $\rho_c$ is a conserved density, $\mathbf{J}_c$ is the associated conserved current, and
\(c\) designates mass, momentum, and energy.
\autoref{eq:cont_eq} follows directly from symmetry arguments: invariance under time
translation yields energy conservation, spatial translation invariances implies
momentum conservation, and an underlying global phase symmetry provides
mass conservation. When \autoref{eq:cont_eq} is defined with periodic boundary conditions, the
volume integrals of the conserved densities remain exact invariants for all time. Thus, both
the local continuity relations and their associated global constraints must be respected:
the domain-integrated mass, the two Cartesian components of momentum, and the total energy
may not drift at any time during the rollout. Any surrogate or reduced-order model that
aspires to physical fidelity must therefore honour these integral invariants, in addition to
satisfying the differential conservation laws \autoref{eq:cont_eq}.

\subsection{Integral Invariants}
For definiteness, we show that mass, momentum, and energy are conserved in this system.
To this end, recall that
\begin{subequations}
\begin{align}
    &J_{\text{mass}}^{j} = \rho_{\text{mass}} v^j \\
    &J_{\text{mom}}^{ij} = \rho_{\text{mass}} v^i v^j + p \delta^{ij} \\
    &J_{\text{energy}}^{j} = (\rho_{\text{energy}} + p)v^j
\end{align}
\end{subequations}
where $p$ is the pressure and $\delta_{ij}$ is the Kronecker delta.
Having introduced pressure as a fifth dynamical variable, a constitutive relation
is needed in order to arrive at a closed system of equations,
which is achieved on writing the energy density as
\begin{equation}
    \rho_{\text{energy}}=\rho_{\text{mass}} e+\frac{1}{2}\rho_{\text{mass}}\lvert\mathbf v\rvert^{2},
\end{equation}
where the specific internal energy $e$ is related to the pressure,
and using the ideal-gas law
\begin{equation}
  p=(\gamma-1)\,\rho_{\text{mass}} e,
\end{equation}
with $\gamma = 1.4$ the adiabatic index of a diatomic gas.

While the continuity equations control the pointwise evolution of
the densities, global conservation guarantees that the total
amount of each conserved quantity is invariant under the flow map.
Integrating the continuity equation for any density
\(\rho_{c}\) over the periodic domain \(\Omega\) and applying
integration by parts (or, equivalently, the divergence theorem) yields
\begin{equation}
  \frac{\mathrm{d}}{\mathrm{d}t}\!\int_{\Omega}\rho_{c}\,
  \mathrm{d}\mathbf A
  \;+\;
  \underbrace{\int_{\partial\Omega}
              \mathbf J_{c}\!\cdot\!\mathbf n\,
              \mathrm{d}S}_{\displaystyle 
              \text{vanishes for periodic }\Omega}=0.
\end{equation}
Therefore,
\begin{equation}
  Q_{c}(t)=\int_{\Omega}\rho_{c}\,\mathrm{d}\mathbf A
\end{equation}
is an integral of motion. This statement precludes secular drift of mass, momentum, or
energy in long simulations, and serves as a primary evaluation metric for surrogate models.
Note that the Navier-Stokes equations can also be expressed in the form \autoref{eq:cont_eq}
and therefore also admit mass, momentum, and energy as conserved variables. However, in
order to sensibly  train a neural network on both Compressible Euler and Navier-Stokes, one
should attach to the model input an indication of whether or not $J_{\text{mom}}$ contains a
viscous term.

\section{Supplementary Figures}
\subsection{Hat Matrix}
\begin{figure}[h] 
    \centering
    \includegraphics[width=0.475\textwidth]{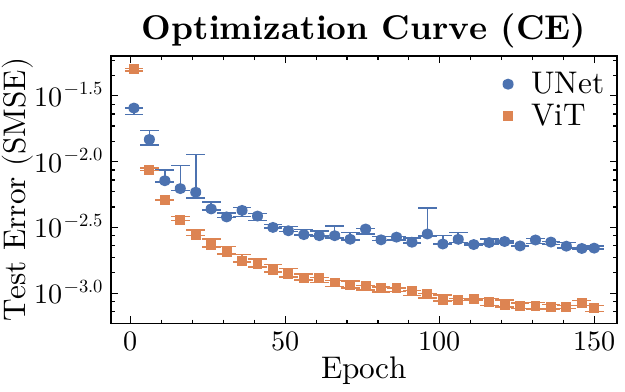} 
\caption{
Optimization trajectories for the CE task, shown as
    test scaled mean squared error (SMSE) versus training epoch for our UNets and our
ViTs. Markers denote the median performance, while vertical
range bars indicate the variance across model seeds at each epoch. Training is
halted in the late-stage learning regime to test whether the induced local gradient
geometry is sufficient to identify a physically consistent, generalizing
solution.
}
    \label{fig:opt_curve_CE_app}
\end{figure}
\begin{figure}[h] 
    \centering
    \includegraphics[width=0.475\textwidth]{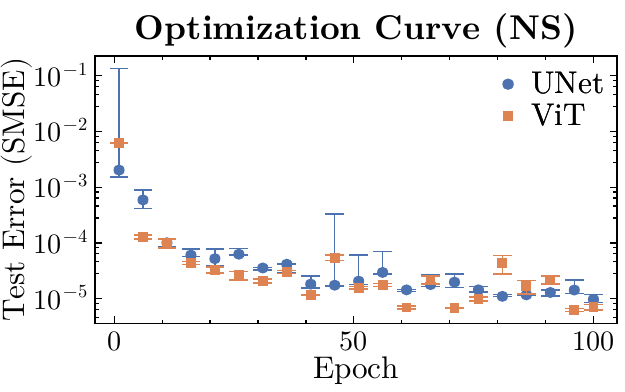} 
\caption{
Same as \autoref{fig:opt_curve_CE_app}, except for NS data. In contrast to the
CE problem, which involves four fields and has greater variability in time, one-step NS
emulation is learned to comparable proficiency by both models.
For representative rollout predictions, see \autoref{fig:rollout_ViT_NS}
and \autoref{fig:rollout_UNet_NS}.
}
    \label{fig:opt_curve_NS}
\end{figure}

\begin{figure}[h] 
    \centering
    \includegraphics[width=0.475\textwidth]{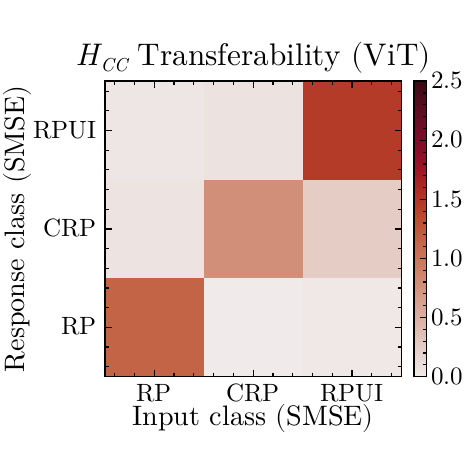}
	\caption{
    Class-to-class transferability matrix for our ViTs trained on the three-class CE split
    labeled RP, CRP, and RPUI. Each entry reports the time-averaged response  of a given
    class (vertical axis)  produced by test example example gradients from the input class
    (horizontal axis). Diagonal dominance indicates class-locked gradient geometry;
    substantial off-diagonal values would imply reuse of dynamical features across classes.
	}
    \label{fig:diag_VC_app}
\end{figure}
\begin{figure}[h] 
    \centering
    \includegraphics[width=0.475\textwidth]{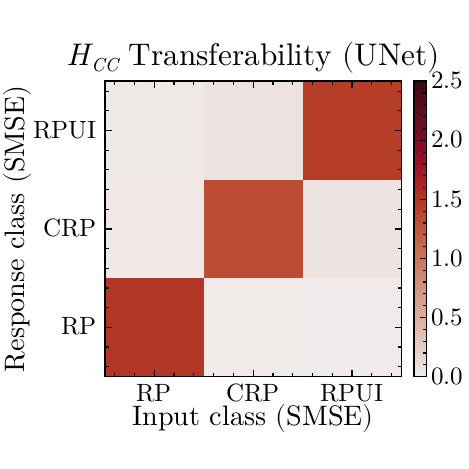}
	\caption{
	Class-to-class transferability matrix for our UNets trained on the CE split, with the same interpretation as \autoref{fig:diag_VC_app}.
    }
    \label{fig:diag_UC}
\end{figure}
\begin{figure}[h] 
    \centering
    \includegraphics[width=0.475\textwidth]{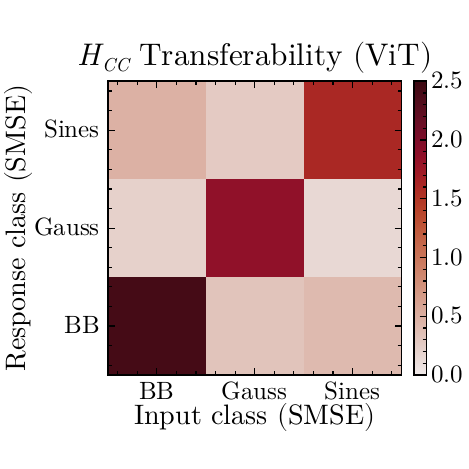}
	\caption{Same as \autoref{fig:diag_VC_app} except for NS data.
	Class-to-class transferability matrix for our UNets under the alternate
	three-class NS split labeled BB, Gauss, and Sines. Each cell is a
	time-averaged influence from one class to another,
	summarizing whether the learned representation supports feature sharing
	between qualitatively distinct initial-condition families. Relatively weak 
	off-diagonal structure indicates that training organizes the
	data into largely disjoint gradient sectors.
	}
    \label{fig:diag_VN}
\end{figure}
\begin{figure}[h] 
    \centering
    \includegraphics[width=0.475\textwidth]{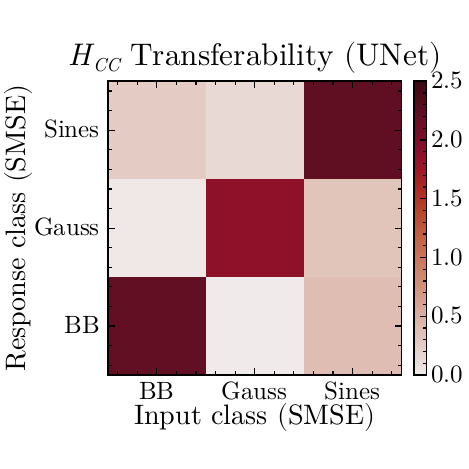}
	\caption{Same as \autoref{fig:diag_VN} except for out UNets.
	}
    \label{fig:diag_UN}
\end{figure}

\begin{figure}[h] 
    \centering
    \includegraphics[width=0.475\textwidth]{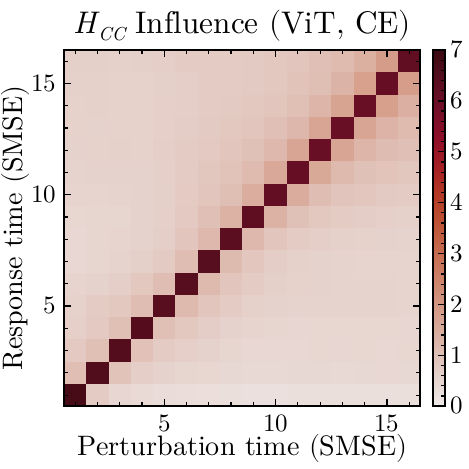}
	\caption{
        Heatmap of two-time influence for our ViTs trained on  CE data,
        shown as a function of perturbation time (horizontal axis) and response time
        (vertical axis). Each pixel reports the intra-class averaged response
        induced by test example gradients at the perturbation time
        and the resulting change in test loss at the response time. A narrow diagonal ridge
        rorresponds to time-local sensitivity consistent with interpolation, rather than
        generalization; substantial off-diagonal structure would indicate time-transferable
        learning. 	}
    \label{fig:heatmap_VC_app}
\end{figure}
\begin{figure}[h] 
    \centering
    \includegraphics[width=0.475\textwidth]{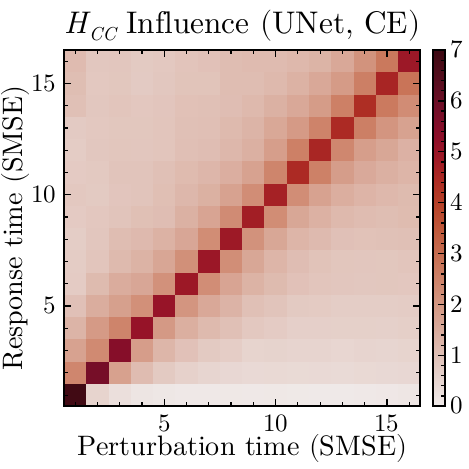}
	\caption{
	Same as \autoref{fig:heatmap_VC_app}, except for our UNets. Desired off-diagonal
	coherence would indicate genuine temporal generalization.}
    \label{fig:heatmap_UC}
\end{figure}
\begin{figure}[h] 
    \centering
    \includegraphics[width=0.475\textwidth]{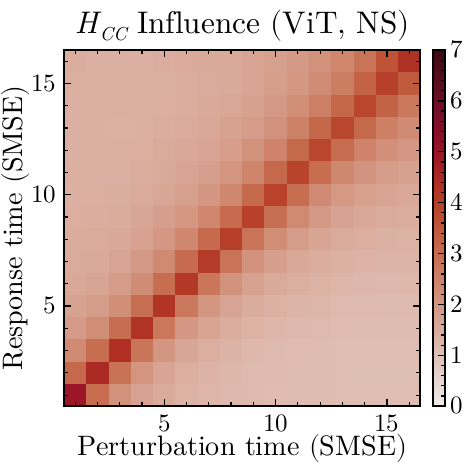}
\caption{
    Same as \autoref{fig:heatmap_VC_app}, except for NS data. The structure measures how
    test-time gradient information at one stage of a rollout
    affects loss at another stage. 
}
    \label{fig:heatmap_VN}
\end{figure}
\begin{figure}[h] 
    \centering
    \includegraphics[width=0.475\textwidth]{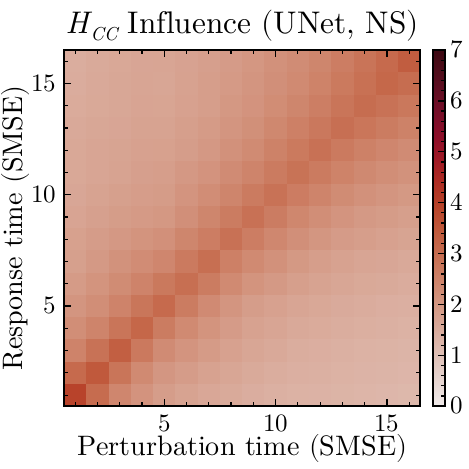}
\caption{
    Same as \autoref{fig:heatmap_VN}, except for our UNets. The enhanced temporal coherence
    relative to compressible Euler is consistent with viscosity-regularized dynamics.
}
    \label{fig:heatmap_UN}
\end{figure}

\begin{figure}[h] 
    \centering
    \includegraphics[width=0.475\textwidth]{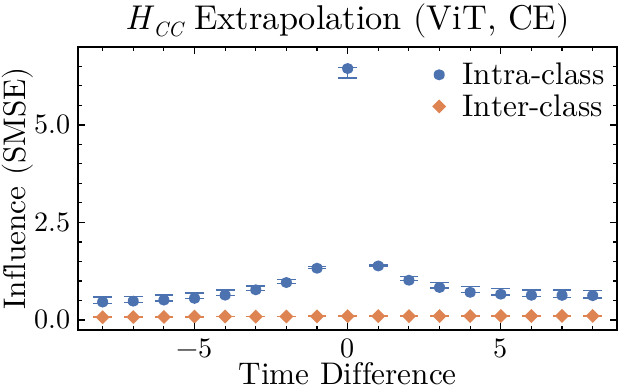}
	\caption{
	Time-lag summary of temporal transferability for our ViTs on CE data. The
	influence is averaged over all time-pairs  of the same
	time difference and then split into intra-class pairs (gradient and
	response drawn from the same initial-condition class) and inter-class pairs
	(distinct initial-condition classes). Strong concentration near zero
	time difference indicates that gradient information fails to propagate
	coherently across time, while the lack off inter-class influence
	reveals an absence of physics consistent generalization. 	
}
    \label{fig:horizon_VC_app}
\end{figure}
\begin{figure}[h] 
    \centering
    \includegraphics[width=0.475\textwidth]{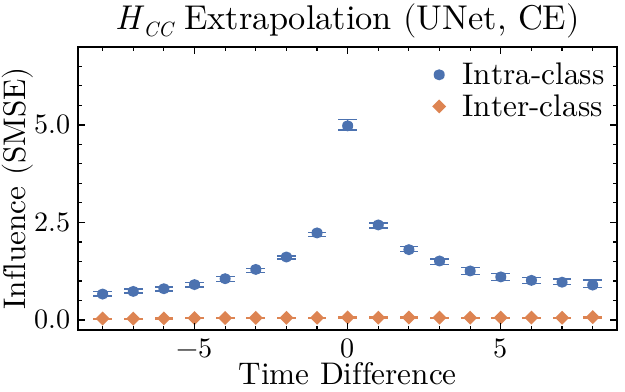}
	\caption{
	Same as \autoref{fig:horizon_VC_app}, except for our UNets.
	The gap between curves demonstrates that temporal coherence is predominantly a
	within-class phenomenon; the model does not learn to share dynamical structure
	across classes.
}
    \label{fig:horizon_UC}
\end{figure}
\begin{figure}[h] 
    \centering
    \includegraphics[width=0.475\textwidth]{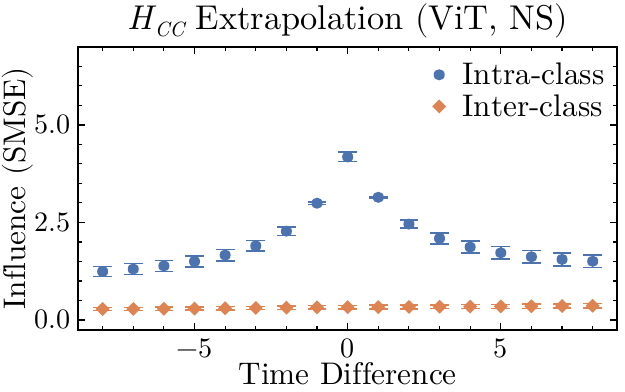}
	\caption{
	Same as \autoref{fig:horizon_VC_app}, except for NS data.
	The intra-class and inter-class averages separate temporal coherence within an
	initial-condition family from cross-family reuse. Persistent suppression of the
	inter-class curve indicates that the model behaves as a collection of
	class-conditioned local estimators rather than a shared dynamical emulator.
	}
    \label{fig:horizon_VN}
\end{figure}
\begin{figure}[h] 
    \centering
    \includegraphics[width=0.475\textwidth]{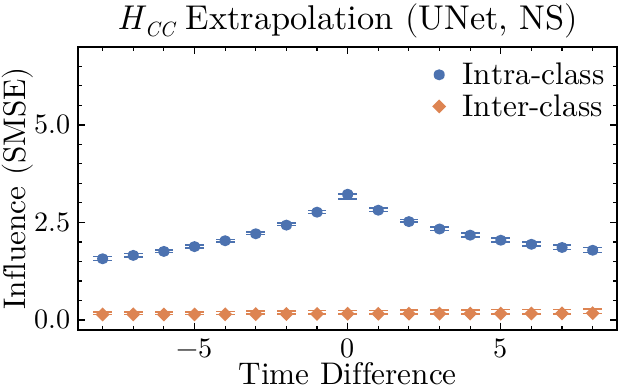}
	\caption{
	Same as \autoref{fig:horizon_VN}, except for our UNets.
	}
    \label{fig:horizon_UN}
\end{figure}

\begin{figure}[h] 
    \centering
    \includegraphics[width=0.475\textwidth]{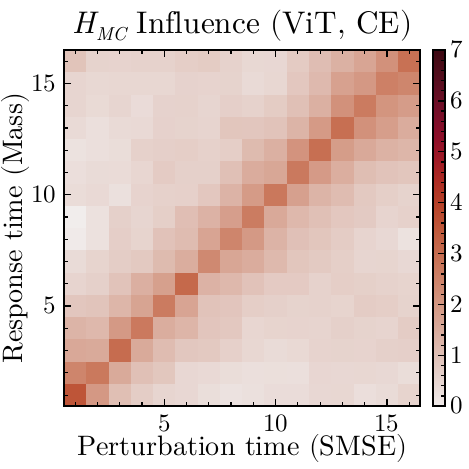}
	\caption{
	Two-time influence map for our ViTs on CE data when the
	response observable is the global mass-consistency signal. The horizontal
	axis indexes the time at which a test perturbation is applied, and the
	vertical axis indexes the time at which the mass-based response is
	evaluated. Off-diagonal support indicates that intra-class mass-related
	gradient information couples distant times.
	}
    \label{fig:heatmap_VC_mass_app}
\end{figure}
\begin{figure}[h] 
    \centering
    \includegraphics[width=0.475\textwidth]{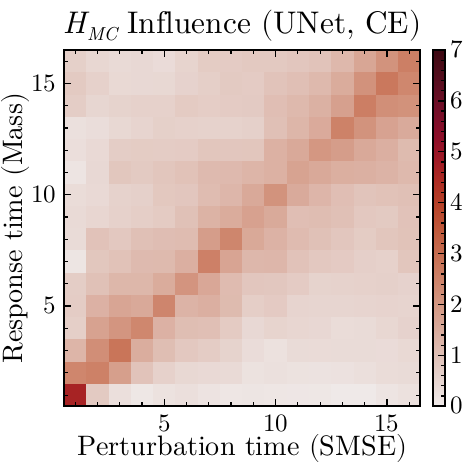}
	\caption{
    Same as \autoref{fig:heatmap_VC_mass}, except for our UNets.}
    \label{fig:heatmap_UC_mass}
\end{figure}
\begin{figure}[h] 
    \centering
    \includegraphics[width=0.475\textwidth]{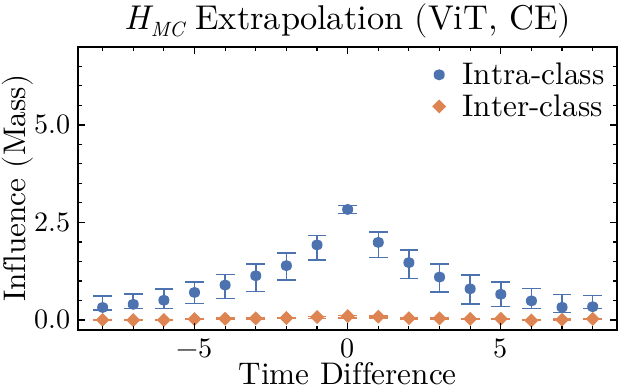}
	\caption{
	Time-lag transferability curve for our ViTs on CE data using the global
	mass-consistency observable. Intra-class averages measure how mass-based
	influence response align within an initial-condition class, while
	inter-class averages measure cross-class reuse of mass-related gradient
	directions. The decay pattern diagnoses whether conservation constraints
	induce transferable structure or remain class-locked.
    }
    \label{fig:horizon_VC_mass_app}
\end{figure}
\begin{figure}[h] 
    \centering
    \includegraphics[width=0.475\textwidth]{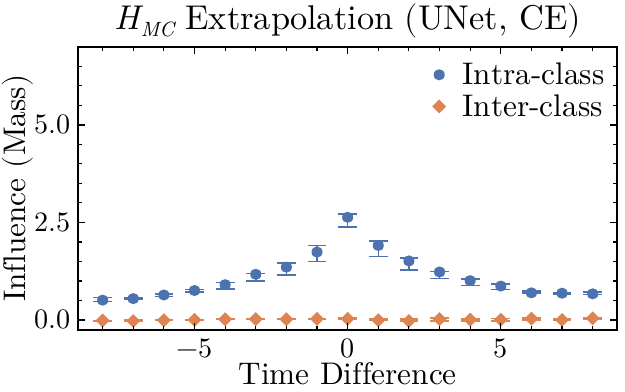}
	\caption{
        Same as \autoref{fig:horizon_VC_mass}, except for our UNets. The intra-class and
        inter-class separation indicates that mass-informed gradients decouple different
        classes, and shared physics learning does not transfer out of class.
	}
    \label{fig:horizon_UC_mass}
\end{figure}

\begin{figure}[h] 
    \centering
    \includegraphics[width=0.475\textwidth]{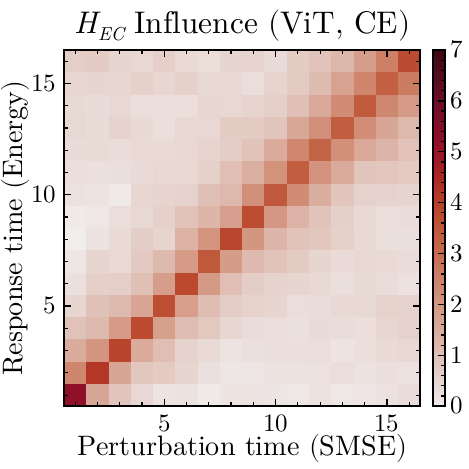}
	\caption{
	Two-time influence map for our ViTs on CE data when the
	response observable is the global energy-consistency signal. The horizontal
	axis indexes the time at which a test perturbation is applied, and the
	vertical axis indexes the time at which the energy-based response is
	evaluated. Off-diagonal support indicates that intra-class energy-related
	gradient information couples distant times.
	}
    \label{fig:heatmap_VC_energy}
\end{figure}
\begin{figure}[h] 
    \centering
    \includegraphics[width=0.475\textwidth]{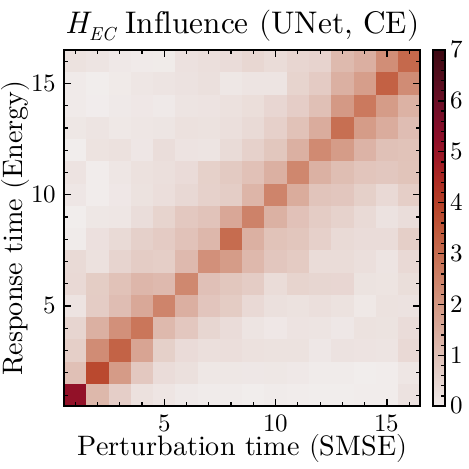}
	\caption{
	Same as \autoref{fig:heatmap_VC_energy}, except for our UNets.
		}
    \label{fig:heatmap_UC_energy}
\end{figure}

\begin{figure}[h] 
    \centering
    \includegraphics[width=0.475\textwidth]{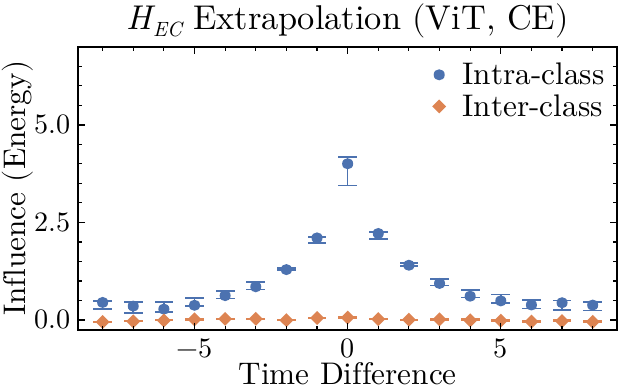}
	\caption{
	Time-lag transferability curve for our ViTs on CE data using the global
	energy-consistency observable. Intra-class averages measure how energy-based
	influence response align within an initial-condition class, while
	inter-class averages measure cross-class reuse of energy-related gradient
	directions. The decay pattern diagnoses whether conservation constraints
	induce transferable structure or remain class-locked.
    }
    \label{fig:horizon_VC_energy}
\end{figure}
\begin{figure}[h] 
    \centering
    \includegraphics[width=0.475\textwidth]{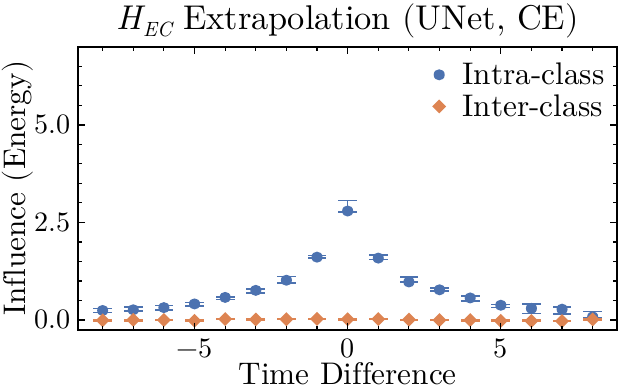}
	\caption{
		Same as \autoref{fig:horizon_VC_energy}, except for our UNets. The
		intra-class and inter-class separation indices energy-informed gradients
		decouple different classes, and shared physics learning does not
		transfer out of class.
	}
    \label{fig:horizon_UC_energy}
\end{figure}

\begin{figure}[h] 
    \centering
    \includegraphics[width=0.475\textwidth]{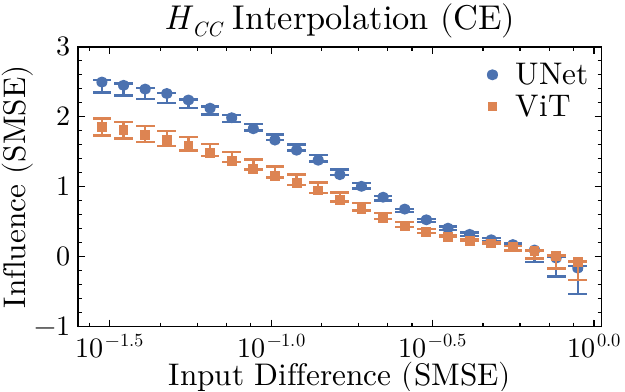}
	\caption{
	Curve fit of the influence as a function of
	feature-space separation between input states for CE data, comparing UNet
	and ViT, rangebars show uncertainty across seeds. Decay to zero indicates
	short-range locality on the learned data manifold, implying that parameter
	updates affect only nearby states and generalization is limited.
	}
    \label{fig:rkhs_CE_app}
\end{figure}
\begin{figure}[h] 
    \centering
    \includegraphics[width=0.475\textwidth]{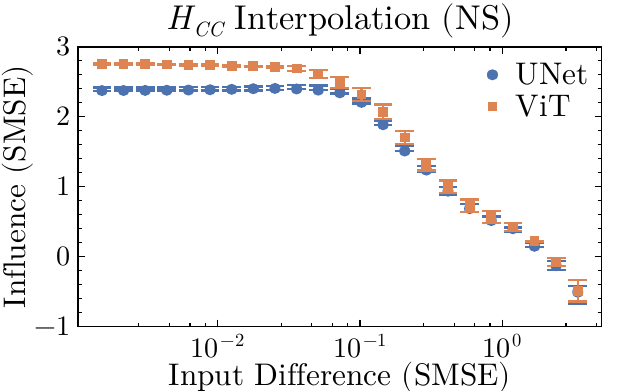}
	\caption{
        Same as \autoref{fig:rkhs_CE}, except for NS data.
	}
    \label{fig:rkhs_NS}
\end{figure}

\begin{figure}[h] 
    \centering
    \includegraphics[width=0.475\textwidth]{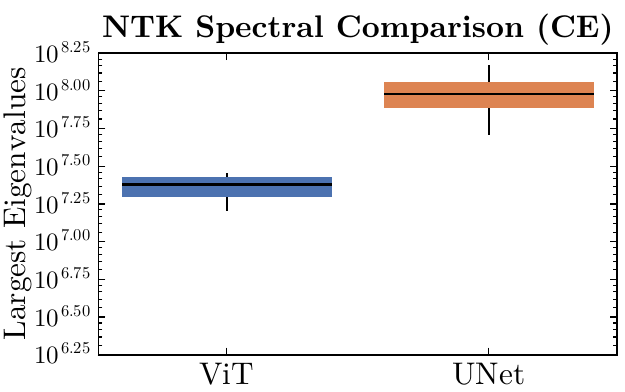}
	\caption{
	Spectral comparison of dominant neural tangent kernel eigenvalues for our UNets
	and our ViTs on CE data. The plotted
	distributions summarize the leading eigenvalue statistics across seeds and
	batches of the trained models. Larger dominant eigenvalues indicate a locally
	stiffer, sharper response geometry.
	}
    \label{fig:eig_CE_app}
\end{figure}
\subsection{Rollout Predictions}
\begin{figure}[h] 
    \centering
    \includegraphics[width=0.475\textwidth]{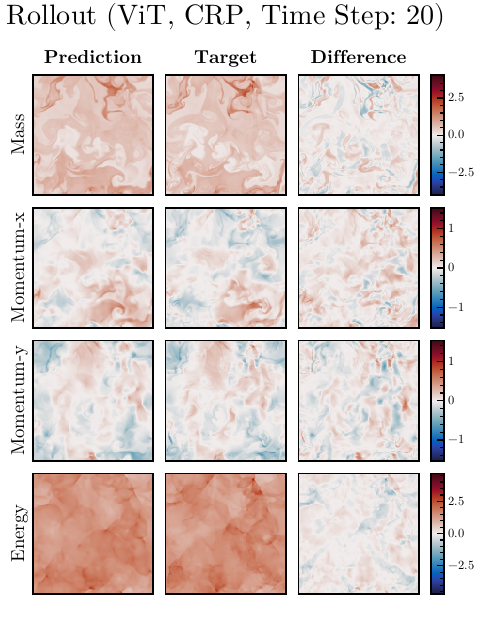}
	\caption{
        Representative final time rollout prediction for one of our ViT models on a CE-CRP initial condition.
	}
    \label{fig:rollout_ViT_CE}
\end{figure}
\begin{figure}[h] 
    \centering
    \includegraphics[width=0.475\textwidth]{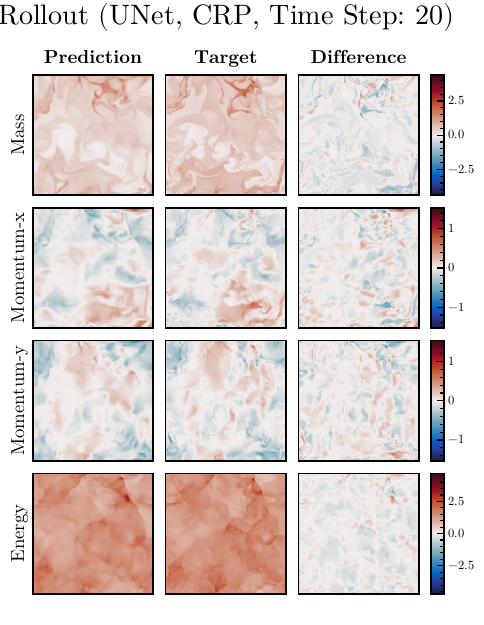}
	\caption{
        Same as \autoref{fig:rollout_ViT_CE} except for one of our UNets.
	}
    \label{fig:rollout_UNet_CE}
\end{figure}
\begin{figure}[h] 
    \centering
    \includegraphics[width=0.475\textwidth]{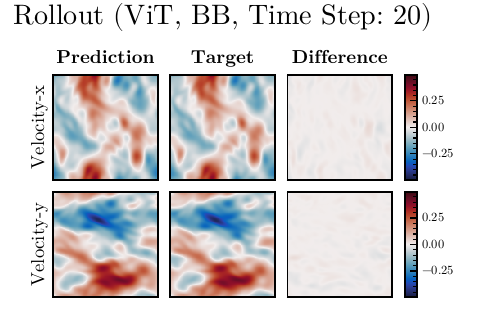}
	\caption{
        Representative final time rollout prediction for one of our ViT models on a NS-BB initial condition.
	}
    \label{fig:rollout_ViT_NS}
\end{figure}
\begin{figure}[h] 
    \centering
    \includegraphics[width=0.475\textwidth]{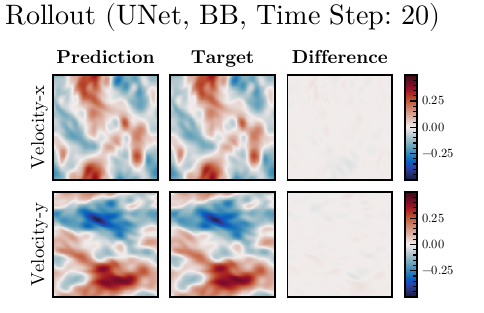}
	\caption{
        Same as \autoref{fig:rollout_ViT_NS} except for one of our UNets.
	}
    \label{fig:rollout_UNet_NS}
\end{figure}


\end{document}